\documentstyle[12pt]{article}

\def\mypagenumber{1}
\def\mydate{June 1, 1995}

\def\myend{\end{document}}

\normalsize

\newcounter{sxn}

\newcounter{axn}

\date{}

\newdimen\mybaselineskip
\newcommand{\beeq}{\begin{equation}}
\newcommand{\eneq}{\end{equation}}
\newcommand{\beqn}{\begin{eqnarray}}
\newcommand{\eeqn}{\end{eqnarray}}

\def\d{\partial}
\def\la{\raise.16ex\hbox{$\langle$} \, }
\def\ra{\, \raise.16ex\hbox{$\rangle$} }
\def\go{\rightarrow}

\def\onehalf{ \hbox{${1\over 2}$} }
\def\psibar{ \psi \kern-.65em\raise.6em\hbox{$-$} }

\def\L{ {\cal L} }
\def\ep{\epsilon}
\def\eps{\varepsilon^{\mu\nu\rho}}

\def\N{\kappa}

\def\vp{ {\vec p} }

\def\ZZ{ \hbox{\tenss Z} \kern-.4em \hbox{\tenss Z} }

\def\L{ {\cal L} }

\def\E{ {\cal E} }
\def\O{{\rm O}}

\def\tr{ {\,{\rm tr}\,} }
\def\Tr{ {\,{\rm Tr}\,} }

\def\tot{ {\rm tot} }

\def\fluct{{\rm fluct}}
\def\RPA{{\rm RPA}}
\def\heavy{{\rm heavy}}
\def\light{{\rm light}}
\def\mass{{\rm mass}}

\def\lave{ {l_{\rm ave}} }
\def\slave{ {l_{\rm ave}^{\,2} } }
\def\ilave{ {l_{\rm ave}^{-1}} }

\begin{document}

\bibliographystyle{unsrt}
\footskip 1.0cm
\thispagestyle{empty}
\setcounter{page}{\mypagenumber}

{\baselineskip=10pt \parindent=0pt \small
\hfill \hbox{\vtop{\hsize=4.5cm UMN-TH-1331/95 \\
hep-th/9505113\\   \mydate ~~(corrected)\\ To appear in Phys.\ Lett.\ B}}
}

\vskip 3cm

\centerline {\LARGE\bf  Spontaneous Magnetization}
\vspace*{5mm}
\centerline {\LARGE\bf  in Lorentz Invariant Theories}
\vspace*{30mm}
\centerline {\large  Denne Wesolowski and  Yutaka Hosotani}

\vspace*{5mm}
\centerline {\small\it School of Physics and Astronomy, University
       of  Minnesota}
\centerline {\small\it Minneapolis, Minnesota 55455, U.S.A.}
\vspace*{5mm}

\centerline {\small\tt wesolowski@mnhep.hep.umn.edu}
\centerline {\small\tt yutaka@mnhep.hep.umn.edu}
\bigskip
\vspace*{25mm}
\normalsize

\begin{abstract}
\baselineskip=17pt

In a  class of three-dimensional Abelian gauge theories
with both light and heavy fermions,  heavy chiral fermions can trigger
dynamical generation of a magnetic field, leading to the spontaneous
breaking of the Lorentz invaiance.  Finite masses of
light fermions tend to restore the Lorentz invariance.
\end{abstract}


\newpage

\baselineskip=22pt


It has been shown in refs.\ \cite{Hoso1,Hoso2} that in a variety of
classes of three-dimensional gauge theories described by
\beqn
\L &=&- {1\over 4} \, F_{\mu\nu}F^{\mu\nu} - {\N_0\over 2} \,
\eps A_\mu \d_\nu A_\rho   \cr
\noalign{\kern 4pt}
&&+ \sum_a  {1\over 2} \, \big[ \,\psibar_a \, , \,
  \big( \gamma^\mu_a (i \d_\mu + q_a A_\mu)
    - m_a \big) \psi_a \, \big] ~,   \label{model}
\eeqn
the Lorentz invariance of the vacuum is spontaneously broken via dynamical
generation of a magnetic field. In other words,  it has been established that
for these models there exists a variational ground state, in which $\la
F_{12}(x)\ra = -B \neq 0,$ that has a lower energy density than the
perturbative ground state.  Quantum fluctuations play a crucial role in
lowering the energy density for these states.\footnote{Cea has considered a
variational ground state with $B\not=0$, which, however, differs from
ours.\cite{Cea}}    In the previous papers models with $\kappa_0 \not= 0$ and
$m_a \ll q_a^2$ were considered.

There are two crucial ingredients.   First, once a magnetic field is
dynamically generated, the fermion energy spectrum is characterized by that of
Landau levels.  The lowest level has an energy $\ep_{a0}=m_a$.  In the
massless fermion limit $\ep_{a0}=0$ so that a perturbative ground state is
infinitely degenerate. This degeneracy is lifted by interactions.
With  certain conditions satisfied,  a variational state, in which the
lowest Landau levels of chirality $+$ fermions are completely occupied,
whereas those of chirality
$-$ fermions are empty, has the lowest energy.\cite{Hoso1}  Secondly it is
recognized that the lowering of the energy density by a non-vanishing
magnetic field is mainly due to  the shift in zero-point energies of photons.
If $\kappa_0\not=0$, a photon is topologically massive in perturbation
theory.  Once
$B\not=0$ is generated, a photon becomes a gapless mode, serving as the
Nambu-Goldstone boson associated with the spontaneous breaking of the
Lorentz invariance.  Hence zero-point energies are shifted, resulting in
lowering the energy density of the ground state.\cite{Hoso2,Hoso3}

Recently Gusynin, Miransky, and Shovkovy have shown that in a model
with an external magnetic field and an additional four-fermi interaction,
flavor-nonsinglet condensate $\la \psibar_a\psi_a\ra$ can be
spontaneously generated.\cite{Miransky,Parwani}  Again the existence
of infinitely degenerate perturbative ground states in the massless fermion
limit plays a crucial role.

We here stay within renormalizable theories described by (\ref{model})
with general  fermion content
to obtain a criterion for  the spontaneous magnetization
($B\not=0$) to occur. We shall also evaluate contributions to the energy
density due to finite fermion masses.

A brief review is in order. In (1) $\psi_a$ is a two-component Dirac spinor.
These spinors, in $2+1$ dimensions, are characterized by the signature  of
the two-dimensional Dirac matrices via the chirality $\eta_a = {i\over 2}
\, \Tr \gamma_a^0\gamma_a^1\gamma_a^2 =  \pm 1$. We utilize the
representation $\gamma^\mu_a = (\eta_a \sigma_3, i \sigma_1, i\sigma_2)$.
Since the model is invariant under charge conjugation and  the
transformation  $m_a \go - m_a$ is  equivalent to the transformation
$\gamma^\mu_a \go - \gamma^\mu_a$  or $\eta_a \go -\eta_a$, one can take
$q_a>0$ and $m_a\ge 0$ without loss  of generality.

As in \cite{Hoso1} we have the expression for the energy density
\beqn
\Delta \E &=& \E(B\not=0) - \E(B=0) \cr
&=& {1\over 2} B^2  + \Delta \E_{\rm f.z.e.} +  \Delta \E_\fluct\cr
\noalign{\kern 10pt}
\Delta \E_\fluct &\sim& \Delta \E_\RPA  \cr
\noalign{\kern 4pt}
&=&-{i \over 2}  \int {d^3p\over (2\pi)^3} ~  \Big\{ \,
 \tr  \ln \,( 1 - \Gamma^{(2)} D_0) \big|_{qB,\nu} - (B \go 0) \, \Big\} \cr
\noalign{\kern 6pt}
&=&-{i \over 2}  \int {d^3p\over (2\pi)^3} ~
\ln { (1+\Pi_0) \bigg\{ 1 + \displaystyle {1\over p^2} (p_0^2 \Pi_0
   - {\vp\,}^2 \Pi_2 ) \bigg\}    - {1\over p^2} (\N_0 - \Pi_1)^2 \over
    (B\go 0 ) }~~.
  \label{energy1}
\eeqn
$\Delta \E_{\rm f.z.e.}$ is the shift in fermion zero-point energies in a
given uniform magnetic field.
$\Delta\E_\fluct$ represents the shift in the energy density due to quantum
fluctuations of fields, and is approximated by the RPA correction
$\Delta \E_\RPA$.   The vertex functions $\Gamma$, the bare and full photon
propagators  $D_0$ and $D$, and the invariant functions  $\Pi$'s are
related by
\beqn
\Gamma^{\mu\nu}(p) &=& (D_0^{-1} - D^{-1})^{\mu\nu} \cr
&=&  ~ (p^\mu p^\nu - p^2 g^{\mu\nu} ) \Pi_0
 + i \ep^{\mu\nu\rho} p_\rho \Pi_1 \cr
&& +(1-\delta^{\mu 0} ) (1-\delta^{\nu 0})
  (p^\mu p^\nu - {\vec p\,}^2 \delta^{\mu\nu}) (\Pi_2 - \Pi_0) .
\eeqn
The superscript for $\Gamma$ in (\ref{energy1}) indicates that
$\Gamma (p)$ is approximated by $\O (e^2)$ diagrams.

 Much as in \cite{Hoso1,Hoso2} we obtain a consistency condition
which the fermion content must satisfy in order that there be symmetry
breaking.  Equations of motion imply,  in the true vacuum,
that $\kappa_0 \la F_{12}\ra = \la J^0 \ra$.   Since
$\la J^0 \ra =  \Pi_1(0) \la F_{12}\ra$,  the relation
$\kappa_0 = \Pi_1(0)$ must be satisfied in order for $B=- \la F_{12}\ra$
to be nonvanishing.   This consistency condition was shown  to
relate to the Nambu-Goldstone theorem associated with the spontaneous
breaking of the Lorentz invariance. The dynamically generated nonzero magnetic
field must accompany a gapless excitation, which is the
photon of the theory.\footnote{The possibility of regarding a photon as a
Nambu-Goldstone boson has been discussed by Bjorken\cite{Bj},
 by Nambu\cite{Nambu}, and by Kovner and  Rosenstein\cite{Kovner}.  Our
picture is different from theirs.  In our case the Lorentz invariance is
spontaneously broken in the physical content.}  Since the induced
Chern-Simons term has a coefficient $-\Pi_1(0)$, one can phrase that
 the bare Chern-Simons term must be exactly cancelled
by the induced Chern-Simons term in order to have B$\neq 0.$

We now consider some definite cases of fermion content.
Suppose that there are light fermions ($m_{\rm light} \ll q^2$) and
heavy fermions ($m_{\rm heavy} \gg q^2$).  In perturbation theory
\beqn
&&\Pi_0 \big|_{B=0}  =  \Pi_2 \big|_{B=0}
=\sum {q_a^2\over 8\pi} {1\over (-p^2)^{1/2} } \,
   \bigg( \sqrt{z_a}  + (1-z_a)   \sin^{-1} {1\over \sqrt{1+z_a}} \bigg)  \cr
\noalign{\kern 5pt}
&&\Pi_1 (p) \big|_{B=0} =
-\sum \eta_a {q_a^2\over 4\pi}  \sqrt{z_a} \,
  \sin^{-1} {1\over \sqrt{1+z_a}} ~, ~~~ z_a = {4m_a^2\over -p^2}~.
  \label{perturbPiOne}
\eeqn
This gives, for the part of heavy fermion
contributions,\cite{Redlich,Ishikawa,Zeitlin}
\beeq
\Pi_1^\heavy (0) \big|_{B=0} = -\sum_\heavy \eta_a {q_a^2\over 4\pi}
 = \kappa_\heavy  ~.
   \label{heavyCS1}
\eneq

In the presence of $B\not= 0$ the energy spectrum of fermions has the
structure of Landau levels  $\ep_{an}= \sqrt{ m_a^2 + 2nq_a|B|}$ ($n=0,1,2,
\cdots$). There is asymmetry in the $n=0$ modes.  $\Pi_1(0)$ is found to be
\beeq
\Pi_1(0) \Big|_{B\not=0} =
{1\over 2\pi}\sum \eta_a q_a^2 (\nu_a - {1\over 2})~.
\label{PiOneB}
\eneq
Here $\nu_a$ is the filling fraction of the Lowest Landau level.  We suppose
that either $\nu_a=0$ (empty) or $\nu_a=1$ (completely filled).

The lowest Landau level has an energy
$m_a$.   Hence for heavy fermions we may assume that the lowest
Landau levels are completely empty, and  assign a filling factor
$\nu_a=0$.   The approximation is valid for $|B| \ll m_\heavy^{3/2}$.
Comparing (\ref{PiOneB}) with (\ref{heavyCS1}), one finds
\beeq
\Pi_1^\heavy(0)\Big|_{B\not=0} = \Pi_1^\heavy(0) \big|_{B=0}
   \equiv - \kappa_\heavy
\eneq
Further heavy fermion contributions to $\Pi_0$ and $\Pi_2$ at low
energies ($|p^2| \ll m^2_\heavy$) are O($q^2/m_\heavy$).
In other words, the only effect of heavy fermions at low energies is
to shift the bare Chern-Simons coefficient $\kappa_0$  to
$\kappa_0+ \kappa_\heavy$.

The variational ground state (true vacuum) is specified with
$\{ \nu_a ~;~ a\in {\rm light~fermions} \} $.
The consistency condition to be satisfied is thus
\beeq
\kappa_0 +\kappa_\heavy =
 {1\over 2\pi}\sum_{\light} \eta_a q_a^2 (\nu_a - {1\over 2}) ~~.
    \label{consistency2}
\eneq
Even if the bare Chern-Simons coeffiecient $\kappa_0=0$,
 the requisite consistency condition can be satisfied.

To be more definite, we consider the following models:

\noindent [Model A]  \quad
The light fermion content is
chirally symmetric.  In other words, light fermions come in a pair
of $(m_a,q_a, \eta_a=\pm )$.     $\{ \nu_a \}$ must satisfy
(\ref{consistency2}).

\noindent [Model B] \quad  This is a special case of model A, in which
 $\nu_a=1$ or 0 for $\eta_a=+$ or $-$, respectively.
In this case the consistency condition is reduced to
\beeq
\kappa_0 +\kappa_\heavy =
 {1\over 4\pi}\sum_{\light}  q_a^2  ~~.
    \label{consistency3}
\eneq

\noindent [Model C] \quad
In model B we suppose that $q_a=q$ and  $m_a=m$ ($a=1 \sim N_f$).
The consistency condition is
\beeq
\kappa_0 +\kappa_\heavy =
 {N_f  q^2 \over 4\pi}  ~~.
    \label{consistency4}
\eneq

The shift in zero-point energies of
photons between the perturbative vacuum and the variational ground state is
given by \cite{Hoso2}
\beqn
\Delta \E &=& \int {d\vp\over (2\pi)^2} ~ {1\over 2}
\Big\{ \omega(\vp\,)_{B\not=0} - \omega(\vp\,)_{B=0} \Big\}  \cr
\noalign{\kern 10pt}
&\sim& \int {d\vp\over (2\pi)^2} ~ {1\over 2}
\bigg\{ \sqrt{ {\vp\,}^2 + \kappa_{tot}^2(p)_{B\not=0}} -
 \sqrt{ {\vp\,}^2 + \kappa_\tot^2(p)_{B=0}}  \,  \bigg\}~,
   \label{energy2}
\eeqn
where $\kappa_\tot(p) = \kappa_0 - \Pi_1(p)$.  In model A
$\Pi_1(p)^\light=0$ identically to O($e^2$).  Hence, for $p^2 \ll m_\heavy^2$,
$\kappa_\tot(p)_{B=0} \sim \kappa_0 + \kappa_\heavy$.
On the other hand the consistency condition
implies that $\kappa_\tot(0)_{B\not=0} =0$.  For large $|\vp\,|$,
$\kappa_\tot(p)_{B\not=0} \sim \kappa_\tot(p)_{B=0}$.   The deviation
develops below $\ilave$ where $\lave$
 is a characteristic magnetic length.  For $m_\light=0$ \cite{Hoso2}
\beqn
{1\over \slave} &=& {\vert \sum_\light \eta_a\nu_a q_a^3\vert \over
{\vert \sum_\light \eta_a\nu_a q_a^2 \vert}} \times \vert B\vert
  \hskip 1cm \hbox{(model A)} \cr
\noalign{\kern 8pt}
&=& \hskip .5cm  {\sum_\light q_a^3 \over \sum_\light q_a^2} \times |B|
 \hskip 1.6cm \hbox{(model B)} ~.
\eeqn

To get an estimate of (\ref{energy2}), we approximate
 $\kappa_\tot(p)_{B\not=0} =0$ and
$\kappa_\tot(p)_{B=0} = \kappa_0+\kappa_\heavy$ for $|\vp\,|\le \ilave$,
whereas   $\kappa_\tot(p)_{B\not=0}
=\kappa_\tot(p)_{B=0}$ for $|\vp\,|\ge \ilave$.  Then
\beqn
\Delta \E &\sim& -{1\over 8\pi} {\vert\kappa_0 +
\kappa_\heavy \vert \over \slave } + \O(l_{ave}^{-3}) \cr
\noalign{\kern 6pt}
&=& - {1\over 32\pi^2} \sum_\light q_a^3 \cdot |B| \hskip 1cm \hbox{(model B)}
   \label{energy3}
\eeqn
for $|\kappa_0 +\kappa_\heavy| \, \lave \gg 1 $.
As (\ref{energy2}) and (\ref{energy3}) are negative definite,  the energy
density
$\E$ is indeed lowered for the vacuum state with $B\neq 0$ provided
that $\kappa_0+\kappa_\heavy $, $\sum_\light \eta_a \nu_a q_a^2$,
and $\sum_\light \eta_a \nu_a q_a^3$ are all non-vanishing and that
the consistency condition (\ref{consistency2}) is satisfied.
In particular, even if the bare Chern-Simons term is absent
($\kappa_0=0$),  heavy fermions can induce a dynamical magnetic field
$B\not= 0$.

The same conclusion is obtained more convincingly by evaluating
(\ref{energy1}). We  recall \cite{Hoso1}
\beqn
&&\Pi_0(p)\big|_{B\not=0} - \Pi_0(p)\big|_{B=0} \cr
\noalign{\kern 4pt}
&&\hskip 1.5cm =  \sum_\light \nu_a { q_a^2\over 2\pi} {1\over p_0} \,
 \Big\{ {1\over 2-p^2 l_a^2 - 2m_a p_0 l_a^2}
   - {1\over 2-p^2 l_a^2 + 2m_a p_0 l_a^2} \Big\} + {\rm O}(B^2) \cr
\noalign{\kern 10pt}
&&\Pi_1(p)\big|_{B\not=0} - \Pi_1(p)\big|_{B=0} \cr
\noalign{\kern 4pt}
&&\hskip 1.5cm = \sum_\light \eta_a \nu_a \, {q_a^2\over 2\pi}  \,
  \Big\{ {1\over 2-p^2 l_a^2 - 2m_a p_0 l_a^2}
   + {1\over 2-p^2 l_a^2 + 2m_a p_0 l_a^2} \Big\} + {\rm O}(B^2)  \cr
\noalign{\kern 10pt}
&&\Pi_2(p) |_{B\not=0} - \Pi_2(p) |_{B=0}
= {\rm O}(B^2)
   \label{approximatePi}
\eeqn
where $l_a^{-2}= |q_a B |$.
When $m_a=0$, $\Pi_0(p)_{B=0}=\Pi_2(p)_{B=0}=
(\sum_\light  q_a^2 /16) (-p^2)^{-1/2}$ and $\Pi_1(p)_{B=0} =
-\kappa_\heavy$ for $|-p^2| \ll m_\heavy^2$.   Inserting these with
(\ref{approximatePi}) into (\ref{energy2}) gives the result \cite{Hoso2}
\beqn
\Delta\E_\RPA &=&
- {\sum_\light \eta_a \nu_a q_a^3 \over 2\pi^3}  \cdot
\tan^{-1} {8 \sum_\light \eta_a \nu_a q_a^2 \over \pi \sum_\light q_a^2}
  \cdot |B|   + \O(|B|^{3/2})   \cr
\noalign{\kern 6pt}
&=&  - {1 \over 4\pi^3}  \tan^{-1} {4 \over \pi } ~\sum_\light  q_a^3
  \cdot |B|   + \O(|B|^{3/2})  \hskip 1cm (\hbox{model B}).
    \label{energy4}
\eeqn
Comparing this with (\ref{energy3}), one finds that the
shift in zero-point energies accounts for  half of the RPA effect.

Recently Cangemi, D'Hoker and Dunne have evaluated the effective potential
in a  magnetic field, supposing that all Landau levels are
empty in the vacuum.\cite{Cangemi}   They found that  a
uniform constant magnetic field configuration shows instability against
forming inhomogeneity.   Cangemi et al.'s variational vacuum state corresponds
to $\{ \nu_a=0 \}$ whereas in our variational vacuum $\nu_a=1$ for $\eta_a=+$.
It is desirable to evaluate the effective potential for a general vacuum state
specified with $\{ \nu_a \}$.

In passing, if $\nu_a=0$ in model A, the consistency condition
(\ref{consistency2}) becomes
$\kappa_0+\kappa_\heavy = -(4\pi)^{-1}\sum \eta_a q_a^2 = 0$.
Further, from the symmetry $\Pi_1(p)_\light=0$.
Hence both $\kappa_\tot(p)_{B=0}$ and
$\kappa_\tot(p)_{B\not=0}$ vanish for $p^2 \ll m_\heavy^2$.
In other words the state with $(B\not=0, \nu_a=0)$  has a higher energy
density than the state with $B=0$.

It is straightforward to evaluate  O($m_a$) corrections to $\Delta \E$.
We evaluate them in model C, supposing $q_a=q, m_a=m \not= 0$.
$\Delta \E/ q^6$ is a function of two dimensionless parameters $m /q^2$ and
$|B|/q^3$.  We shall find an expression for $\Delta \E$ valid for
$m^2/q^4 \ll |B|/q^3 \ll 1$.  (Note $(m l)^2 = m^2/ (q|B|) \ll 1$.)

Corrections  to $\Delta \E_\RPA$ are found from (\ref{energy1}) and
(\ref{approximatePi}).
$\Pi_0$ and $\Pi_1$ give corrections of O($m$) and O($m^2$), respectively.
Hence a dominant correction comes from $\Pi_0$ and is found to be
\beqn
\Delta \E_\RPA^{\rm mass} &=&
{N_f^2 q^4 m \over 48\pi^3} \, f(y) \qquad {\rm where} ~~
      y=\Big( {16\over N_f} \Big)^2 {|B|\over q^3}
  \hskip 1cm (\hbox{model C})    \cr
\noalign{\kern 14pt}
f(y) &=& y \int_0^\infty dk \, {k^3 (\sqrt{y}\, k+1) \over
                 (\sqrt{y}\, k+1)^2 (k^2+2)^2 + (16 k^4 /\pi^2) } \cr
\noalign{\kern 10pt}
&\sim& {1\over 2(1+16/\pi^2)} ~
 y ~ \bigg\{ - \ln y  +1.8 \bigg\}
  \hskip 1cm {\rm for}~~ y < 0.1 ~~.
  \label{massenergy1}
\eeqn
Notice that $\Delta \E_\RPA^{\rm mass} >0$.

Secondly the lowest Landau level has an energy $m_a$ so that a contibution
from occupied levels is
\beeq
\Delta\E_{\rm occupied}^\mass = \sum_\light  {\nu_a m_a  \over 2\pi l_a^2}
 = {1\over 2\pi} \sum_\light \nu_a m_a q_a \, |B| ~.
   \label{massenergy2}
\eneq
It is positive.

There is another important contribution coming from a shift in fermion
zero-point energies.   Landau levels are given by
$\epsilon_{an} = \sqrt{m_a^2+(2n /l_a^2)}$.  Recalling that there is
asymmetry in the lowest level, i.e.\ the $n=0$ level exists for either a
particle or an anti-particle, one finds that fermion zero-point energies in
the presence of  a uniform magnetic field are
\beqn
\Delta\E_{\rm f.z.e.}
&=& \sum_\light {1\over 2\pi l_a^2}
\bigg\{ - {1\over 2} m_a
  - \sum_{n=1}^\infty
\sqrt{m_a^2+{2n\over l_a^2}}  \bigg\}  ~~.
     \label{massenergy3}
\eeqn

To evaluate this sum we utilize the standard zeta-function technique, taking
\beeq
\zeta (s)=\sum_{n=1}^\infty {1\over \lambda_n^s} = \sum_n {1\over \Gamma (s)}
\int_0^\infty dt \, t^{s-1} e^{-\lambda_n t}.
\eneq
Here we have $\lambda_n \equiv m^2 + (2n/l^2) $ and need to calculate $\zeta
(-{1\over 2}).$
The substitution $u\equiv 2nt / l^2 $ immediately leads to
\beqn
\zeta (s) &=& \bigg( {l^2\over 2}\bigg)^s {1\over \Gamma (s)}
\sum_{n=1}^\infty {1\over n^s}
\int_0^\infty du\,  u^{s-1} \exp \Big( -u-{m^2l^2\over 2n}u \Big) \cr
&=& \bigg( {l^2\over 2} \bigg)^s \zeta_R(s)
- m^2 \bigg( {l^2\over 2} \bigg)^{s+1} \zeta_R(s+1) + \cdots~~.
\eeqn
Here $\zeta_R$ is the  Riemann zeta function and the expansion is valid
for $m^2 l^2 \ll 1$.    In particular
\beqn
\zeta(-\onehalf) &=&
- {1\over 4\pi} \zeta_R(\hbox{${3\over 2}$}) \bigg( {l^2\over 2} \bigg)^{-1/2}
 - m^2 \bigg( {l^2\over 2} \bigg)^{1/2} \zeta_R(\onehalf) + \cdots~~.
  \label{ourzeta}
\eeqn
Note $\zeta_R({3\over 2})=2.61238$ and  $\zeta_R(\onehalf)= -1.46035 <0$.

Utilizing (\ref{ourzeta}) in (\ref{massenergy3}), one finds
\beqn
\Delta\E_{\rm f.z.e.}
&=& \sum_\light {1\over 4\sqrt{2} \pi^2} \zeta_R(\hbox{${3\over 2}$})
  {1\over l_a^3} - \sum_\light {m_a \over 4\pi l_a^2} + \O (m^2) \cr
\noalign{\kern 8pt}
&=& {1\over 4\sqrt{2} \pi^2} \zeta_R(\hbox{${3\over2}$})
  \sum_\light q_a^{3/2} |B|^{3/2}
- {1\over 4\pi} \sum_\light m_a q_a  |B|  + \O (m^2) \cr
\noalign{\kern 8pt}
&&\hskip 4.cm {\rm for} ~~ {m^2\over q^4} \ll {|B|\over q^3} \ll 1
   \label{fze}
\eeqn
Notice that the second term (linear in $m_a$) is negative.
This behavior has been noticed by Gusynin,  Miransky, and Shovkovy in
a different context.\cite{Miransky} ~
(\ref{massenergy2}) and  (\ref{fze}) combine to give
\beqn
&&\Delta\E_{\rm occupied}^\mass + \Delta\E_{\rm f.z.e.}  \cr
\noalign{\kern 8pt}
&&\hskip 0cm = {1\over 4\sqrt{2} \pi^2} \zeta_R(\hbox{${3\over2}$})
  \sum_\light q_a^{3/2} |B|^{3/2}
+{1\over 2\pi} \sum_\light (\nu_a - \onehalf) m_a q_a  |B|  + \O (m^2) ~.
  \label{fermionEnergy}
\eeqn
We recognize that in model B the second term vanishes.

Combining the  results of (\ref{energy4}), (\ref{massenergy1}),
 and (\ref{fermionEnergy}), one finds, in model C,
\beqn
\noalign{\kern 8pt}
{1\over q^6}\, \Delta\E &=&  {1\over 2} {B^2\over q^6}
- {N_f\over 4\pi^3}   \tan^{-1} {4  \over \pi} \cdot {|B|\over q^3}
+ {N_f \zeta_R({3\over 2})\over 4\sqrt{2} \pi^2} \cdot
\Big( {|B|\over q^3} \Big)^{3/2} \cr
\noalign{\kern 8pt}
&&\hskip 2cm  + {N_f^2 m\over 48 \pi^3 q^2}\,
   f\Big( {256|B| \over N_f^2 q^3} \Big)  + \O (m^2, B^2) \cr
\noalign{\kern 12pt}
&\sim& {1\over 2} \Big({N_f\over 16}\Big)^2 y^2
- 0.117 \Big({N_f\over 16}\Big)^3 \, y
 + 0.749 \Big({N_f\over 16}\Big)^4 \, y^{3/2} \cr
\noalign{\kern 10pt}
&&\hskip 2cm
+ 0.0328 \Big({N_f\over 16}\Big)^2 \,{m\over q^2} \,y ( -\ln y + 1.8 )
  + \O (m^2) \cr
\noalign{\kern 12pt}
&&\hskip 3cm {\rm for} ~~ {m^2\over q^4} \ll y =
\Big( {16\over N_f} \Big)^2 {|B|\over q^3} < 0.1
 \label{totalenergy}
\eeqn
We have restored the Maxwell energy term $\onehalf B^2$.  Other O($B^2$)
corrections have much smaller coefficients.

We observe that finite masses of light fermions tend to restore the Lorentz
symmetry.  $\Delta\E (y)$ develops a minimum at $y_{\rm min}\not=0$.
As $m$ gets bigger, $y_{\rm min}$ decreases and  $\Delta\E (y_{\rm min})$
increases.  More specifically
\beqn
&& \hbox{
\begin{tabular}{l l l l l }
\noalign{\kern 6pt}
\hline
\noalign{\kern 6pt}
&$m/q^2$  &\hskip .5cm  $y_{\rm min}$ &\hskip .3cm $\Delta\E_{\rm min}/q^6$
         & $(ml)^2 |_{\rm min}$ \\
\noalign{\kern 6pt}
\hline
\noalign{\kern 6pt}
$N_f=16 ~~~$  &0.    &$9.2 \times 10^{-3}$ & $- 3.7 \times 10^{-4}$ &
\\  &0.02  &$8.7 \times 10^{-3}$ & $- 3.3 \times 10^{-4}$
                           &\hskip .4cm  0.05 \\
\noalign{\kern 6pt}
\hline
\noalign{\kern 6pt}
$N_f=~ 4$ &0.    &$1.0 \times 10^{-1}$ & $- 7.1 \times 10^{-5}$ & \\
 &0.02  &$7.2 \times 10^{-2}$ & $- 3.5 \times 10^{-5}$
                          &\hskip .4cm  0.09 \\
\noalign{\kern 6pt}
\hline
\noalign{\kern 6pt}
\end{tabular}
}
\label{table1}
\eeqn

\noindent  For larger values of $m/q^2$ the expressions (\ref{ourzeta}) and
(\ref{totalenergy}) are not accurate.

In this paper we have shown that  heavy fermions can induce spontaneous
magnetization through the induced Chern-Simons coefficient at low energies.
Further we showed that nonvanishing masses of light fermions tend to
restore the Lorentz symmetry by giving a positive slope in the $|B|$
dependence near the minimum of $\Delta\E$.

Our consideration here was limited to the dependence of the energy density
on a dynamically generated magnetic field.  In view of the recent result by
Gusynin et al.\ \cite{Miransky} it is of great interest to find the effective
action as  a function of both dynamically generated magnetic field $B$ and
chiral condensate $\la \psibar_a\psi_a \ra$.  It may be that both $B$ and
$\la \psibar_a\psi_a \ra$ are generated dynamically in a cooperative manner.
We shall come back to this point in the near future.

\bigskip

\noindent {\it Note added:}  Recently Kanemura and Matsushita
have examined the behaviour of the model (\ref{model}) at finite temperature
in the massless fermion limit.\cite{Kanemura}   They have found that
the coefficient of the linear term $\propto |B|$ in the energy density
(\ref{energy4}) or (\ref{totalenergy}) remains negative even at finite
temperature.

\vskip .5cm
\baselineskip=16pt
\centerline{\bf Acknowledgements}
\bigskip

This work was supported in part
by the U.S.\ Department of Energy under contract no. DE-AC02-83ER-40105.
D.W. would like to thank C.L. Erickson and L.D.\ Reed
for many stimulating and useful discussions.

\end{document}